\documentclass[12pt,letterpaper]{article}
\usepackage[dvips]{graphics}
\usepackage{setspace}

\makeatletter

\newcommand{\LyX}{L\kern-.1667em\lower.25em\hbox{Y}\kern-.125emX\spacefactor1000}

\makeatother

\begin{document}

\vspace{-2cm}

\begin{flushright}
UT-Komaba 98-4\\
UTHEP-378\\
DPNU-98-10
\end{flushright}
\begin{center}
{\Large Multicanonical simulation of 3D dynamical triangulation model 
and a new phase structure}
\end{center}
\vspace*{0.5cm}

{\centering{\large Tomohiro Hotta}\large \par}

{\centering \textit{Institute of Physics, University of Tokyo,\\
Meguro-ku, Tokyo 153, Japan}\par}

\vspace{0.2cm}

{\centering {\large Taku Izubuchi}\large \par}

{\centering \textit{Institute of Physics, University of Tsukuba,\\
Tsukuba, Ibaraki 305, Japan}\par}

\vspace{0.1cm}

{\centering {\large and}\large \par}

\vspace{0.1cm}

{\centering {\large Jun Nishimura}\large \par}

{\centering \textit{Department of Physics, Nagoya University, \\
Chikusa-ku, Nagoya 464-01, Japan}\par}
\vspace{0.2cm}

\begin{center}
\large February, 1998
\end{center}

\vspace{0.2cm}

\begin{abstract}
We apply the multicanonical technique
to the three dimensional dynamical triangulation
model, which is known to exhibit a first order phase transition
with the Einstein-Hilbert action.
We first clarify the first order nature of the phase transition with
the Einstein-Hilbert action in several ways including 
a high precision finite size scaling analysis.
We then add a new local term to the action and confirm the conjecture
made through the MCRG technique
that the line of the first order phase transition
extends to the expanded phase diagram, ending at a point.
Fractal dimension at the end point is measured to be around three 
up to the present size.
\end{abstract}

\newpage

\section{Introduction}

Quantum gravity is one of the most important topics in particle physics.
The attempts to construct it within ordinary field theories
confront a difficulty due to the fact that the theory is
perturbatively unrenormalizable.
This does not mean immediately that the attempts fail, 
since it might be possible to construct it nonperturbatively
as is the case with the 3D nonlinear sigma model.
The lattice regularization provides a natural approach in this context.
In the case of quantum gravity, since 
the dynamical variable is given by the metric of the space time,
how to discretize the space time is itself a nontrivial problem.
Above all, it seems difficult to preserve the general coordinate invariance
manifestly.
The success of the dynamical triangulation model 
in two dimensions \cite{BIPZ} summarized
by its equivalence to the Liouville theory \cite{Liouville} 
in the continuum limit 
shows that the model provides a natural framework to study quantum gravity
nonperturbatively within ordinary field theories.
Although the restoration of the general coordinate invariance in the 
continuum limit is only supported by the success in two dimensions, 
we might consider that the summation over all the triangulation
with the equal weight forms the basis of the restoration.
The situation with the other lattice regularization of quantum gravity
known as the Regge calculus is very obscure \cite{Regge}.

The above point of view motivated
intensive studies of the dynamical triangulation 
model in higher dimensions 
through Monte Carlo simulation.
It turned out, however, that with an action obtained by naive discretization
of the Einstein-Hilbert action, the phase transition observed is 
of first order both in three dimensions \cite{3DQG}
and in four dimensions \cite{BBKP96,4DDT1st}.
The signal of the first order phase transition is much weaker in 4D 
than in 3D.
We note that this does not necessarily mean that 
it is easier to construct a continuum theory
out of this model in 4D than in 3D,
and we consider that this is merely 
related to the existence of singular vertices
in the crumpled phase in 4D \cite{Progress,singular}.
By adding a local term which suppresses the singular vertices,
we observe a first order phase transition in 4D, which is even stronger
than in 3D.
Indeed, from the viewpoint of perturbative unrenormalizability,
the situation in 4D must be severer than in 3D.
The $\epsilon$-expansion of $(2+\epsilon)$-dimensional quantum gravity
\cite{twopluseps} suggests the existence of a non-trivial fixed point,
and this result is of course more reliable for smaller $\epsilon$ naively.
We therefore concentrate on 3D dynamical triangulation model in this paper.
Three-dimensional quantum gravity is worth being studied
in itself, since it can be also regarded as 
the first quantization of membranes,
which is considered to be a 
fundamental object in the M-theory \cite{Mtheory}.

In lattice theories in general,
it is often preferable (\textit{e.g.} improved actions to reduce the finite
lattice spacing effects) or even necessary (\textit{e.g.} the Wilson term to
eliminate fermion doublers) 
to modify the lattice actions from that obtained by naive
discretization of the action in the continuum. We may therefore hope to obtain
a second order phase transition in the present case by adding a local term to
the naive Einstein-Hilbert action. 
Once the continuum limit can be taken, the continuum theory constructed
is expected to be independent of the details of the constructions
thanks to universality. 
Adding some matter fields or gauge fields 
\cite{4DMatter} is another possibility
to obtain a continuum theory,
though it changes the theory to be constructed.
We stick to pure gravity as the simplest case in this paper.

We perform a numerical simulation of this model using
the multicanonical technique \cite{MC}, 
which is powerful especially
in the presence of large potential barriers between some local minima,
as in the present case 
in which the system undergoes a first order phase transition. 
We measure 
the quantity which corresponds to the `temperature' in the microcanonical
ensemble,
which enables us to extract various important
thermodynamical properties of the model very clearly.
We first clarify the first order nature of the phase transition with
the Einstein-Hilbert action in several ways including 
a high precision finite size scaling analysis.
We next add a new term to the action, 
which is motivated by a possible contribution 
from the measure for the path integral
over the metric.
The modified action
has been first studied in four dimensions 
\cite{BrugmannMarinari}.
Recently it has been studied in three dimensions 
by the Monte Carlo renormalization group 
and a possibility was suggested that the first order phase transition
ceases to exist at a finite coefficient of the new term \cite{Renken}.
We confirm this possibility by investigating the expanded phase
diagram \footnote{While this paper was being completed,
a preprint appeared \cite{Catterall}, 
in which the authors address the same issue
but with a conclusion which is different from ours.}.

The phase transition is expected to become second order at the end point
of the phase transition line.
We therefore attempt to observe the continuum physics at this end point.
We measure the boundary area distribution, whose counterpart in 2D
is known to show a clear scaling behavior \cite{KKMW,TsudaYukawa}.
We indeed see a reasonable scaling behavior.

The paper is organized as follows.
In Section \ref{Sec:MC}, we explain the multicanoical techniques 
and introduce 
the quantity which corresponds to the `temperature' in the microcanonical
ensemble.
In Section \ref{Sec:Result}, 
we show the numerical results with the Einstein-Hilbert action,
and clarify the first order nature of the phase transition.
Above all,
we observe a clear finite size scaling,
which is consistent with the first order phase transition. 
In Section \ref{Sec:ModAction}, we introduce the model with the new term.
In Section \ref{Sec:EndPoint}, we show the results with
the modified action and confirm that the first order phase transition
line extends to the expanded phase diagram,
ending at a point.
In Section \ref{Sec:FractalStruct},
we show the results for the boundary area distribution
at the end point, which is consistent with fractal dimension three.
Section \ref{Sec:Concl} is devoted to conclusions
and discussions.

\section{Multicanonical technique 
and ``microcanonical inverse temperature''\label{Sec:MC}}

The naive discretization of the Einstein-Hilbert action 
in the dynamical triangulation
model in three dimensions gives \cite{3DQG}
\begin{equation}
\label{eq:S0}
S_{EH}=-\kappa _{0}N_{0}+\kappa _{3}N_{3}\, ,
\end{equation}
where \( N_{0} \) and \( N_{3} \) are the number of vertices and 
3-simplices respectively in the
simplicial manifold. 
$\kappa _{0}$ corresponds to the inverse of the gravitational
constant, while
$\kappa _{3}$ corresponds to the cosmological constant.
Since we consider ensembles with fixed $N_3$,
we use only the first term $S_0= - \kappa _{0}N_{0}$ in the following.

The partition function of the system with fixed $N_3$
is given as follows.
\begin{eqnarray}
Z(\kappa _{0};N_{3}) & \equiv  & \sum _{\{T(N_{3})\}}\exp (-S_{0})
\nonumber \\
 & = & \sum _{N_{0}}P(N_{0},\, \kappa _{0};\, N_{3})\: ,
\label{eq:PartFunc} \\
P(N_{0},\kappa _{0};N_{3}) & \equiv  & n(N_{0};N_{3})\, 
\exp (\kappa _{0}N_{0})\: .\label{eq:SpectFunc} 
\end{eqnarray}
The summation in the first line of (\ref{eq:PartFunc}) 
is taken over all possible triangulations of 3-dimensional
simplicial manifolds with sphere topology
that contain \( N_{3} \) \( 3 \)-simplices. 
\( n(N_{0};N_{3}) \) is the number of the possible triangulations
with a further restriction that they contain \( N_{0} \) vertices. 
$P(N_0,\kappa_0;N_3)$ represents the node number distribution for
a canonical ensemble with fixed $\kappa_0$.
In the following we suppress the argument $N_3$ in the 
$P(N_0,\kappa_0;N_3)$ and $n(N_0;N_3)$, unless necessary.

As in all of the recent numerical works on higher dimensional dynamical
triangulation model,
we use the so-called ($p$,$q$) moves, which are proven to be
ergodic \cite{GrossVarsted}, to simulate this system.
Actually, $N_3$ is not kept fixed in the moves.
We therefore have to allow a small fluctuation of $N_3$ around 
a target $N_3$ with an appropriate Gaussian potential for $N_3$.
Measurement has been done only when the $N_3$ agrees with the 
target $N_3$.
We refer to it as a simulation for fixed $N_3$
for simplicity.

The system is known to undergo a first order phase transition
at some critical $\kappa_0$.
Near the critical point, $P(N_0,\kappa_0)$ exhibits a double-peak
structure.
As one increases the system size $N_3$,
the valley between the double peaks becomes deeper and deeper,
which makes the ordinary canonical simulation increasingly difficult,
due to the long Monte Carlo time required
to travel back and forth between the two peaks for sufficiently many times.
We perform the multicanonical simulation to overcome this problem.
For technical details, we refer the readers to Ref. \cite{Fujitsu}.
The simulation is performed by replacing
the Boltzmann weight $\exp (\kappa_0 N_0)$ 
with some function of $N_0$, which is determined iteratively
so that the distribution of $N_0$ becomes as constant as possible
within some window of $N_0$.
If the window is wide enough to cover the whole support of the 
distribution function of $N_0$ for a canonical ensemble 
with some $\kappa_0$,
one can reproduce the expectation values of any observables
in the canonical ensemble by taking account of the effect of 
replacing the Boltzmann weight in updating the system.
The idea is essentially a generalization of the histogram method.
If one extends the window by fine-tuning the Boltzmann
weight, the simulation includes information of
canonical ensembles with various $\kappa_0$ in some finite region.
This enables, for example, 
a precise identification of the peak of the node number
susceptibility
as a function of $\kappa_0$,
which is essential in the high precision finite size scaling analysis.

\begin{figure}
{\centering
\resizebox*{1.05\textwidth}{!}{\includegraphics{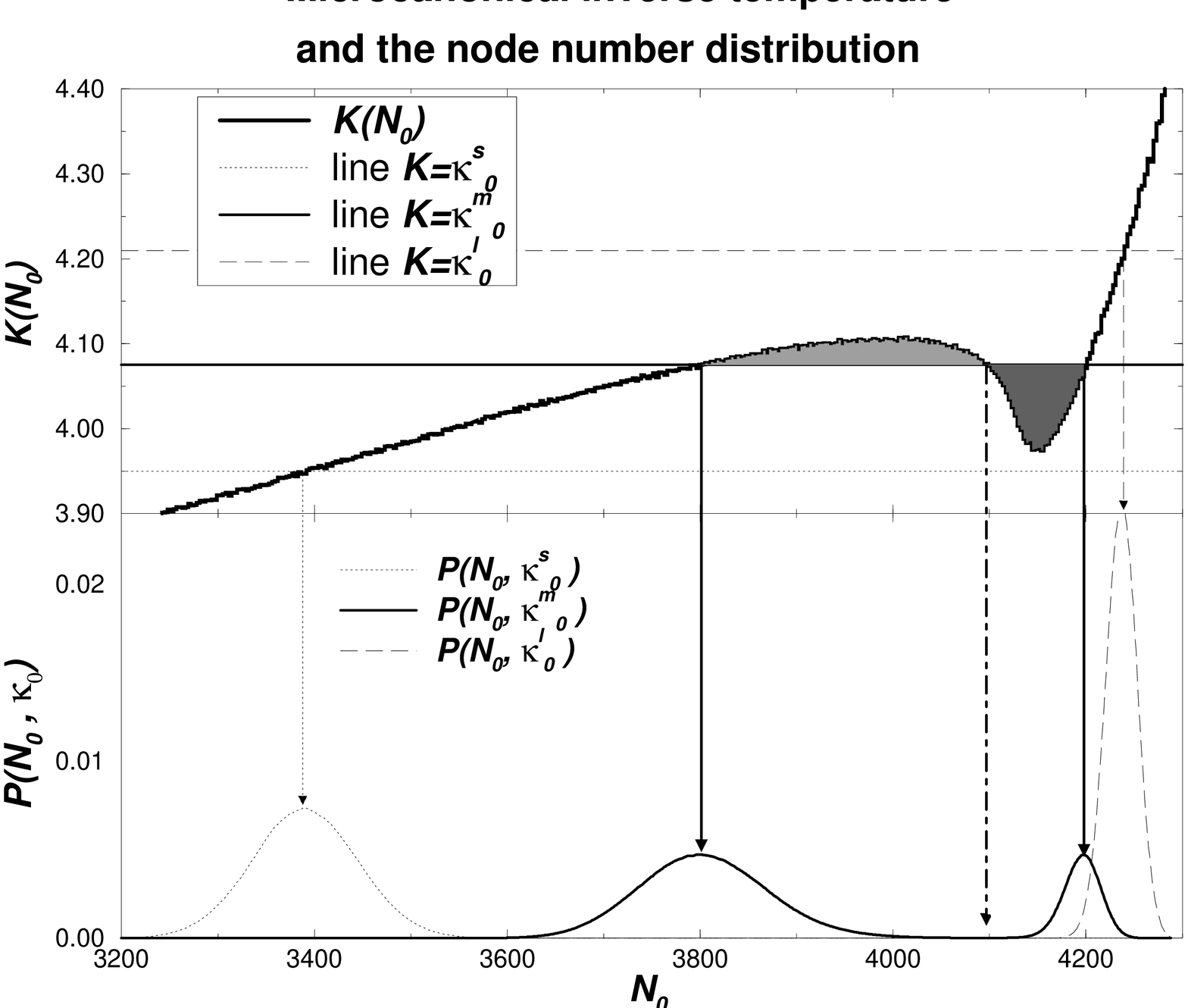}} 
\par}
\caption{The upper part shows the microcanonical inverse temperature 
\protect\( K(N_{0})\protect \) (heavy solid curve)
with three horizontal lines $K=\kappa_0$ corresponding to 
$\kappa_0 =\kappa ^{s}_{0}$, $\kappa ^{m}_{0}$ and $\kappa ^{l}_{0}$.
The lower part shows
the node number distribution
\protect\( P(N_{0},\kappa _{0})\protect \) 
for each of the three values of $\kappa_0$.
\label{fig:Expl}}
\end{figure}

Another advantage of the multicanonical simulation
is that it allows us to extract
the entropy as a function of $N_0$ defined by
\begin{equation}
\mathcal{S}(N_{0})  = \log \, n(N_{0})\, ,\label{eq:EnthropyDef} 
\end{equation}
or its first derivative
\begin{equation}
K(N_{0})  =  - \{ \mathcal{S}(N_{0}+1)-\mathcal{S}(N_{0}) \}.
\label{eq:KappaDef} 
\end{equation}
$K(N_{0})$ is the analogue of the inverse temperature $\beta = 1/T$
defined in the microcanonical ensemble if we regard $-N_0$ as the energy.
We therefore refer to $K(N_{0})$ as 
the microcanonical inverse temperature.
Strictly speaking, the simulation allows us to extract
the entropy $\mathcal{S}(N_{0})$ 
only up to an additional constant independent of $N_0$.
Once $\mathcal{S}(N_{0})$ is known, one can calculate the 
node number distribution
$P(N_0,\kappa_0)$ in the canonical ensemble for arbitrary $\kappa_0$
through
\begin{equation}
\label{eq:SpectRatio}
P(N_0,\kappa_0) = \exp \{ \kappa_0 N_0 + \mathcal{S}(N_{0}) \} .
\end{equation}
From this, we have
\begin{equation}
\label{eq:SpectRatio2}
\log P(N_0 +1, \kappa_0) - \log P(N_0, \kappa_0) 
= \kappa_0 - K(N_0),
\end{equation}
which means that
if \( \kappa _{0}-K(N_{0}) \) is positive (negative) for
some \( N_{0} \), 
the node number distribution \( P(N_{0},\kappa _{0}) \) increases
(decreases) as $N_0$ increases.
This is illustrated in Fig. \ref{fig:Expl}.
The upper part shows the
microcanonical inverse temperature \( K(N_{0}) \) as a function
of $N_0$, with three horizontal lines $K=\kappa_0$ corresponding
to $\kappa_0=\kappa ^{s}_{0}$, $\kappa ^{m}_{0}$ and $\kappa ^{l}_{0}$.
The lower part shows the node number distribution
\( P(N_{0},\kappa _{0}) \)
for each of the three values of \( \kappa _{0} \). 
The intersections of the curve of the microcanonical inverse
temperature and the horizontal lines give the extrema of 
the node number distribution for each $\kappa_0$.

When $\kappa_0=\kappa^{m}_0$,
there are three intersection points: 
$\kappa ^m_{0}=K(N^{(1)}_{0})$, $K(N^{(2)}_{0})$, and
$K(N^{(3)}_{0})$, where $ N^{(1)}_{0} <  N^{(2)}_{0}< N^{(3)}_{0}$.
The node number distribution has the double peaks
at \( N^{(1)}_{0} \) and \( N^{(3)}_{0} \).
\( N^{(2)}_{0} \) gives the place of the local minimum.
From (\ref{eq:SpectRatio2}), 
the ratios
$P(N^{(1)}_0,\kappa^{m}_0)/P(N^{(2)}_0,\kappa^{m}_0)$
and
$P(N^{(3)}_0,\kappa^{m}_0)/P(N^{(2)}_0,\kappa^{m}_0)$
are given by the exponential of the area
of the light and heavy shaded regions in Fig. \ref{fig:Expl}, respectively. 
Since we have chosen $\kappa^{m}_0$ so that
the area of the two regions is equal, the peaks of 
the node number distribution have the same height.

In general,
when the microcanonical inverse temperature has a local minimum
and a local maximum,
the node number distribution shows a double peak structure
for the canonical ensemble
with $\kappa_0$ set to the value between the local minimum and 
the local maximum.
The quantity, therefore, serves as a useful probe to clarify the
fate of the first order phase transition when one modifies
the action from the Einstein-Hilbert action.

\section{The results for the Einstein-Hilbert action
\label{Sec:Result}}
In this section, we show the results for the Einstein-Hilbert action
and clarify the first order nature of the phase transition 
by the measurement of various thermodynamical quantities.

\begin{figure}
{\centering \resizebox*{1.10\textwidth}{!}{\includegraphics{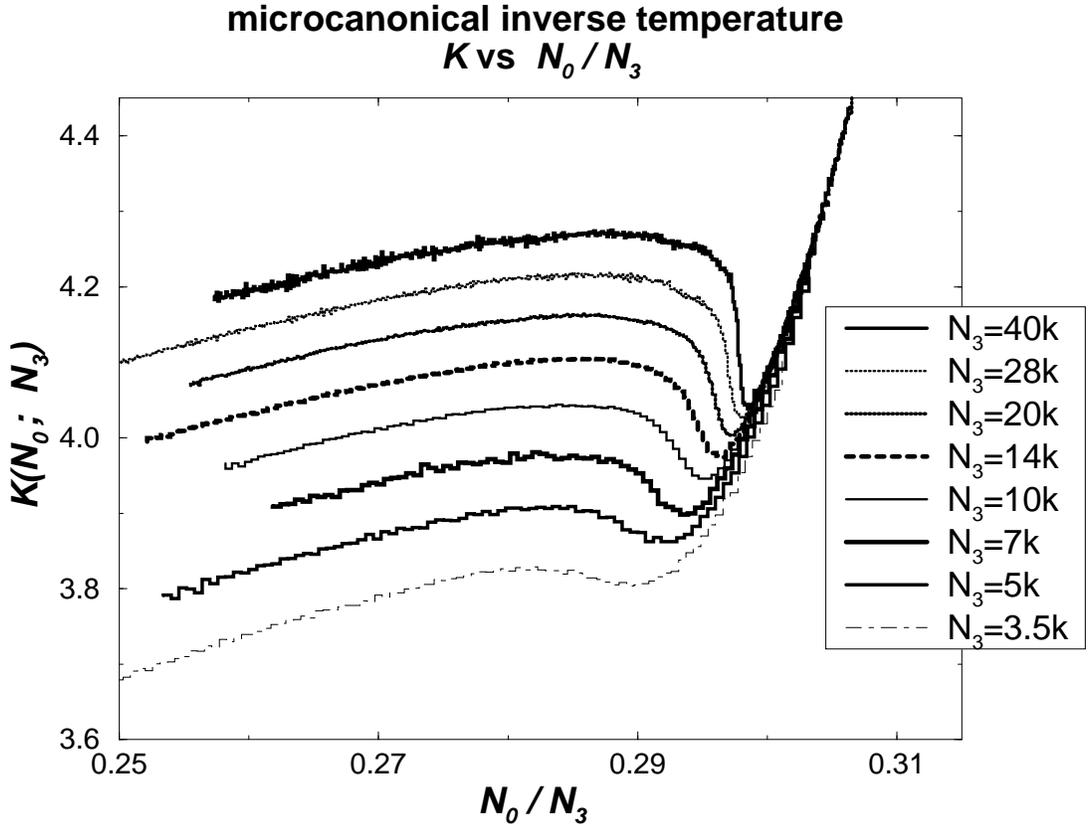}} \par}
\caption{
The microcanonical inverse temperature
\protect\( K(N_{0};N_{3})\protect \) is
plotted against the node number
\protect\( N_{0}\protect \) normalized 
by the system size \protect\( N_{3}\protect \)
for $N_3=3,500$ to 40,000. 
\label{fig:MC-K}}
\end{figure}

\begin{figure}
{\centering \resizebox*{0.8\textwidth}{!}{\includegraphics{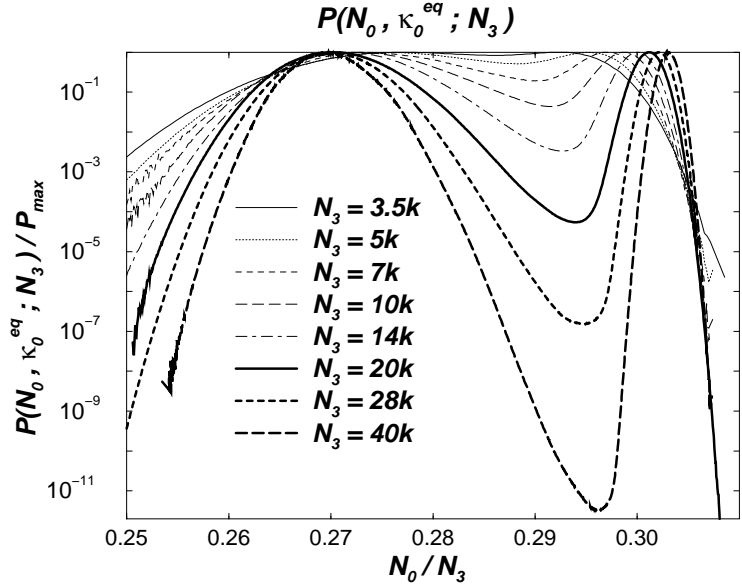}} \par}
\caption{
The node number distribution at $\kappa_0 = \kappa^{eq}_0$
normalized by the peak height, $P_{max}$,
is plotted against $N_0/N_3$ for $N_3=3.5$k, $\cdots 40$k.
\label{fig:MCEquiv}}
\end{figure}

We first show the microcanonical inverse temperature \( K(N_{0};N_{3}) \) 
for various system size from $N_{3}=3,500$ to $40,000$
in Fig. \ref{fig:MC-K}.
Since the microcanonical inverse temperature has a local minimum
and a local maximum,
the node number distribution shows a double peak structure
at some region of $\kappa_0$.
We choose the $\kappa_0$ 
with which the height of the peaks is the same.
We denote such $\kappa_0$ by $\kappa^{eq}_0$.
The determination of $\kappa^{eq}_0$
can be done easily by referring to the microcanonical inverse
temperature as is explained in the previous section.
Fig. \ref{fig:MCEquiv} shows the results for the node number
distribution at $\kappa^{eq}_0$ with the 
system size from $N_3 = 3,500$ to $40,000$.
The logarithm of the ratio 
of the height of the peaks to that of the valley
gives the interfacial energy. (See Ref. \cite{Fujitsu} for further
analysis on this quantity.)
Note that the ratio is about $10^{12}$ for $N_3=40,000$.
One can easily understand that 
ordinary canonical simulations cannot be used
in this situation.

\begin{figure}
{\centering \resizebox*{0.8\textwidth}{!}{\includegraphics{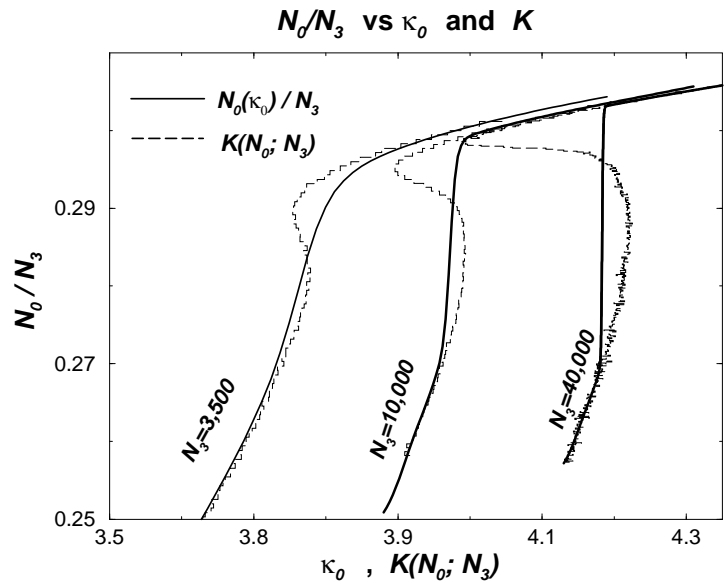}} \par}
\caption{The mean node number normalized by the system size $N_3$
is plotted against $\kappa_0$ by the solid line.
The microcanonical inverse temperature is replotted
by the dashed line for comparison.
 \label{fig:MCcurv}}
\end{figure}

In Fig. \ref{fig:MCcurv}, 
we plot the mean node number in the canonical ensemble
with $\kappa_0$ against $\kappa_0$.
We replot the microcanonical inverse temperature 
with $\kappa_0 = K(N_0;N_3)$ for comparison.
The mean node number shows a sharp increase in the critical region
of \( \kappa _{0} \) for a sufficiently large 
system size,
which is another way of observing the signal of the first order
phase transition.
The data curve of the mean node number coincides with the data curve
of the microcanonical inverse temperature
except in the critical region, as is expected.

\begin{figure}
{\centering \resizebox*{0.8\textwidth}{!}{\includegraphics{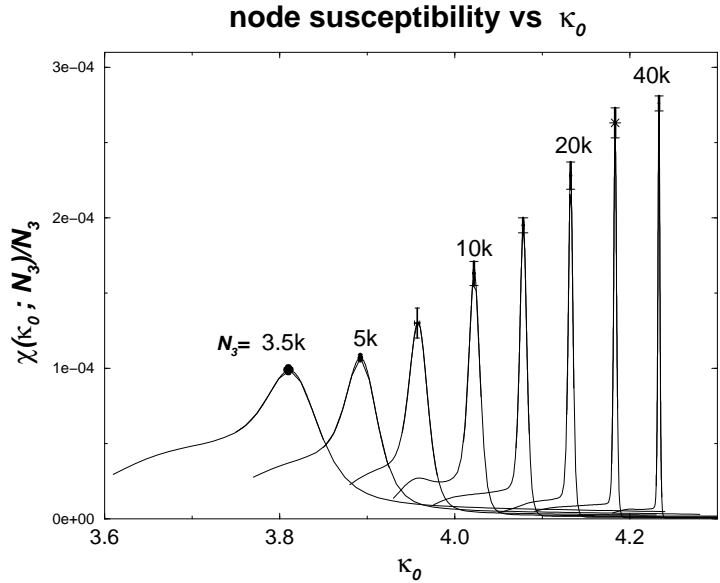}} \par}

\caption{The node number susceptibility 
\protect\( \chi  (\kappa_0; N_3 ) \protect \) 
is plotted against $\kappa_0$
for the system size
from $N_3=3,500$ to $40,000$.
\label{fig:MC-SUS}}
\end{figure}

\begin{figure}
{\centering \resizebox*{1.0\textwidth}{!}{\includegraphics{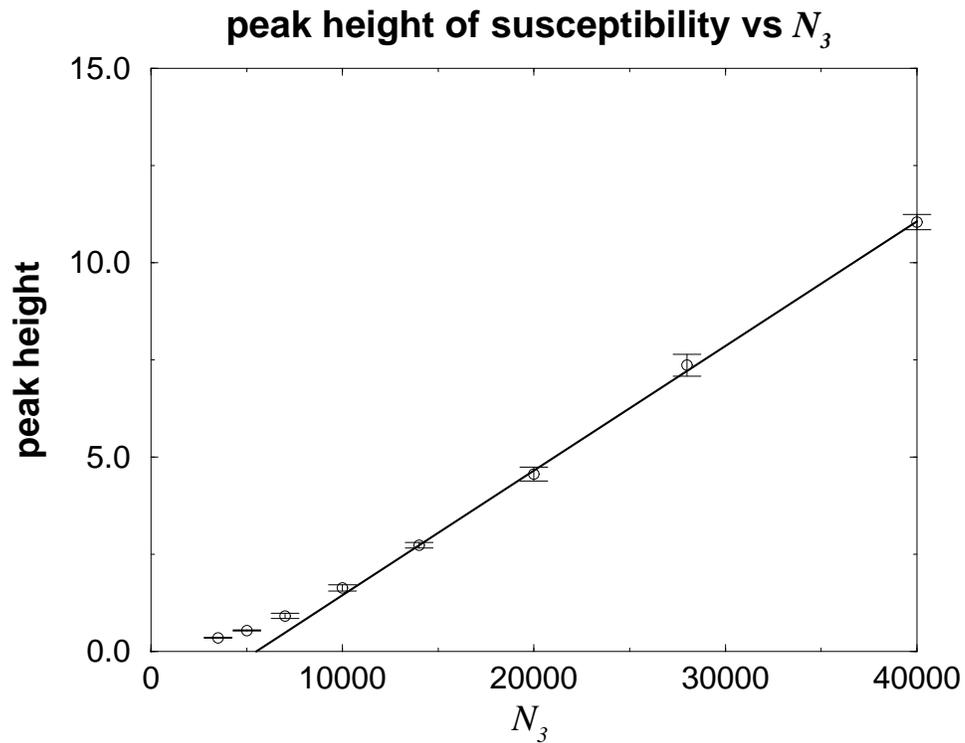}} \par}

\caption{The peak height of the 
susceptibility 
is plotted against $N_3$.
The straight line show the linear fit
$\chi/N_3 = p_1+ p_2 N_3$, 
which is consistent with the first order phase transition.
The $\chi ^2$ of the fit is 
\protect\( \chi ^{2}/dof= 0.3\protect \). 
We use only the data for the four large $N_3$'s.
\label{fig:SUSfit}}
\end{figure}

In Fig. \ref{fig:MC-SUS} we 
show the result for the node number susceptibility 
\( \chi (\kappa_0 ; N_3) \equiv (\left\langle N^{2}_{0}\right\rangle 
-\left\langle N_{0}\right\rangle ^{2})/N_{3} \).
We call the $\kappa_0$ which gives the peak of the susceptibility for
each $N_3$ as the pseudo-critical point
$\kappa^c_0 (N_3)$.
In ordinary statistical systems,
the peak height of the susceptibility 
grows as a power of the system size.
If the peak is due to a first order phase transition,
the internal energy distribution shows a double peak structure
as the one in Fig. \ref{fig:MCEquiv}, and the tunneling
between the two peaks causes a linear growth of the peak height
of the susceptibility.
In Fig. \ref{fig:SUSfit} we plot the peak height of the 
susceptibility against the system size.
The data for the four large $N_3$'s can be fitted very nicely
to the formula
\[
\chi (\kappa ^{c}_{0} (N_3);N_3)=p_{1}+p_{2}N_{3}\, ,\]
with $p_1 = - 1.8(2)$ and $p_2 =3.20(7)\times 10^{-3}$,
which is consistent with 
the finite size scaling for the first order phase transition.

The pseudo-critical coupling constant \( \kappa ^{c}_{0} (N_3) \) 
shifts to the right with no tendency of convergence 
as one increases the system size up to \( N_{3}=40,000 \).
This behavior should be contrasted with 
ordinary statistical systems, for which
the dependence of the pseudo-critical coupling on the system
size \( V \) obeys the scaling relation such as
\begin{equation}
\label{eq:k0FSS}
\kappa ^{c}_{0}(V)=\kappa ^{c}_{0}(\infty ) + c V^{-p}.
\end{equation}
The unusual shift of the pseudo-critical coupling constant 
has been discussed earlier in 
the four-dimensional dynamical triangulation model \cite{jansmit}.

\section{The modified action\label{Sec:ModAction}}
In this section, we modify the Einstein-Hilbert action (\ref{eq:S0})
and attempt a search for a second order phase transition
point in an expanded phase diagram.
We consider the following action.
\begin{eqnarray}
S & = & S_{0}-\mu M\, ,\label{eq:totAction} \\
M & = & \sum _{v}\log \left[ o(v)/4 \right] \, ,\label{eq:MUterm} 
\end{eqnarray}
where \( \mu  \) is the new coupling constant 
\footnote{Note that the $\mu$ defined here is the same as
the one in Refs. \cite{Renken,Catterall}.
The definition of $\mu$ used in Ref. \cite{proceedings}
corresponds to $-\mu$ in this paper.}, and 
\( o(v) \) is the number of \( 3- \)simplices that
contain the node \( v \). The summation in (\ref{eq:MUterm}) is 
taken over all the nodes in the
manifold. 
This term \cite{BrugmannMarinari} can be considered 
as a possible contribution from the path integral
measure \( \Pi _{x}\, g(x)^{\mu /2} \), since \( o(v) \) 
can be interpreted as the ``local volume'', which corresponds
to \( \sqrt{g(x)} \) in the continuum. 

In the following we consider the canonical ensemble 
with fixed \( N_{3} \) given by the partition function : 
\begin{eqnarray}
Z^{(mod)}(\kappa _{0};\mu;N_3 ) 
& = & \sum _{\{T(N_{3})\}}
\exp (\kappa _{0}N_{0}+\mu M)\, ,\label{eq:PartFuncmu0} \\
 & = & \sum _{N_{0}}z(N_{0};\mu;N_3 )\, 
\exp (\kappa _{0}N_{0})\: ,\label{eq:PartFuncmu} 
\end{eqnarray}
where \( z(N_{0};\mu;N_3 ) \) 
is the partition function for fixed \( N_{0} \), which can be given by
\begin{eqnarray}
z(N_{0};\mu ;N_3)=\sum _{\{T(N_{3},N_{0})\}}
\exp (\mu M)\: , & \label{eq:SpectFuncmu} 
\end{eqnarray}
where the summation is now taken over all possible triangulations that contain
\( N_{3} \) \( 3 \)-simplices and \( N_{0} \) vertices. 
As before, we suppress the argument $N_3$ in 
$z(N_{0};\mu ;N_3)$ unless necessary.
In this paper, we concentrate on the negative $\mu$ region,
in which the first order phase transition line is predicted to
end at a point through the MCRG analysis in Ref. \cite{Renken}.

\section{End point of the first order phase transition\label{Sec:EndPoint}}

Let us define 
the (generalized)
microcanonical entropy 
\begin{equation}
\mathcal{S}(N_{0};\mu ) 
\equiv \log z(N_{0};\mu ) 
\end{equation}
and its first derivative
\begin{equation}
K(N_{0};\mu )\equiv 
- \{ \mathcal{S}(N_{0}+1;\mu) -  \mathcal{S}(N_{0};\mu) \},
\end{equation}
which we keep on referring to as
the microcanonical inverse temperature. 

\begin{figure}
{\centering \resizebox*{1.05\columnwidth}{!}{\includegraphics{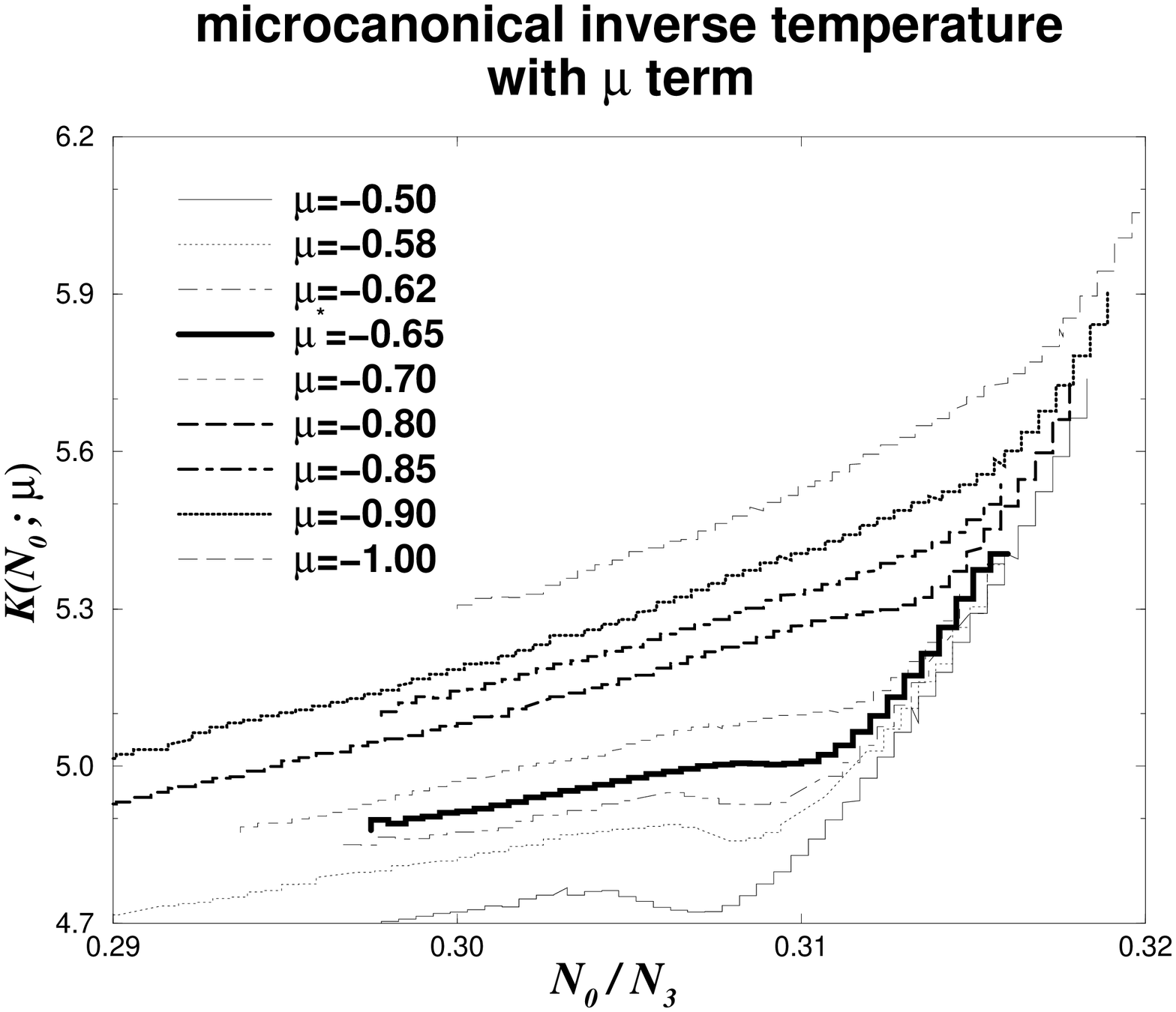}} \par}

\caption{The microcanonical inverse temperature 
\protect\( K(N_{0};\mu )\protect \) as a function of 
\protect\( N_{0}\protect \) for several \protect\( \mu \protect \)'s
with the system size \protect\( N_{3}=10,000\protect \).\label{fig:K}}
\end{figure}

In Fig. \ref{fig:K}, we show the 
microcanonical inverse temperature \( K(N_{0};\mu ) \) 
for several negative values
of \( \mu  \). As was explained in Section \ref{Sec:MC}, the double
peak structure in the node number distribution,
which is a signal of the first order phase transition, can be probed directly
by looking at the microcanonical inverse temperature \( K(N_{0};\mu ) \). 

One can see that the local minimum and the local maximum observed 
for \( \mu =0 \) merge
at \( \mu ^{*}=- 0.65 \), beyond which the \( K(N_{0};\mu ) \) 
becomes a monotonous function of \( N_{0} \). This suggests that
the first order phase transition ceases to exist at \( \mu ^{*} \), 
beyond which it becomes a crossover. 

\begin{figure}
{\centering \resizebox*{0.75\columnwidth}{!}
{\includegraphics{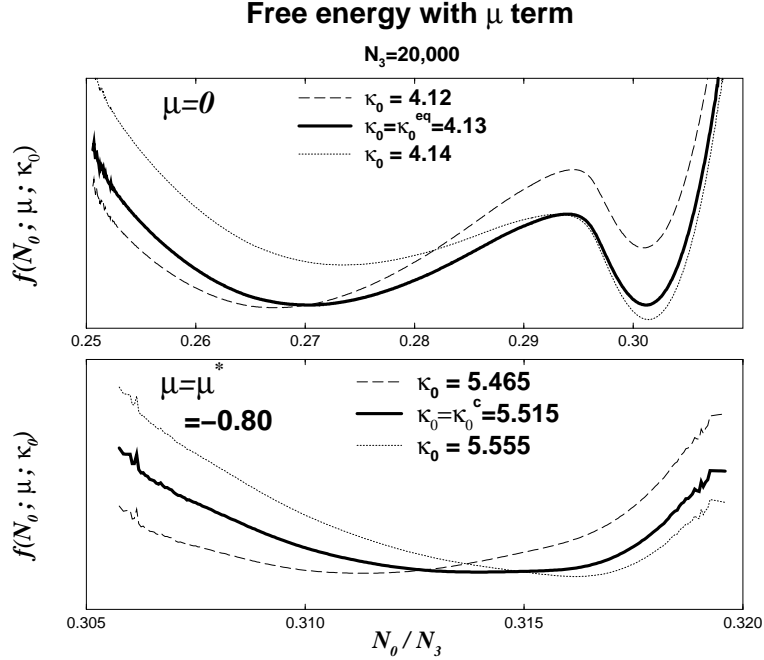}} \par}

\caption{The free energy \protect\( f(N_{0};\, \mu ,\kappa
_{0})\protect \) as a function of \protect\( N_{0}\protect \) near
\protect\( \kappa _{0}^{c}\protect \) for \protect\( \mu =0\protect \)
and 
\protect\( \mu =\mu ^{*} = - 0.80 \protect \) with the system size
\protect\( N_{3}=20,000\protect \).\label{fig:FreeEnergy}}
\end{figure}

For each \( \mu  \ge \mu ^{*} \), 
we define the pseudo-critical coupling 
\( \kappa _{0}^{c} \)
as the position of the peak of the node number
susceptibility \( \chi _{0}(\kappa _{0},\mu ) \). The end point of the
phase transition line is given by 
\( \mu ^{*}=-0.65,\, \kappa ^{c}_{0}=5.00 \) for \( N_{3}=10,000 \), 
and \( \mu ^{*}=-0.80,\, \kappa ^{c}_{0}=5.515 \) for \( N_{3}=20,000 \). 
In Fig. \ref{fig:FreeEnergy}, we plot the
free energy \( f(N_{0};\, \mu ;\kappa _{0})
= - \mathcal{S}(N_{0};\mu )-\kappa _{0}N_{0} \) 
as a function of \( N_{0} \) near \( \kappa _{0}^{c} \) 
for \( \mu =0 \) and \( \mu =\mu ^{*} \). 
The two local minima observed for \( \mu =0 \) merge
into one at \( \mu =\mu ^{*} \). 
The free energy curve at \( (\kappa ^{c}_{0},\mu ^{*}) \) 
has a wide flat bottom, which suggests
the existence of a massless mode. 

\section{Fractal structure at the end point\label{Sec:FractalStruct}}

In order to examine a possible continuum limit at the end point, we measure
the boundary area distribution \( \rho (A,r) \), 
which is the number of boundaries with the
area \( A \) at the geodesic distance \( r \). 
The corresponding quantity in two dimensions
called the loop length distribution is calculated analytically and is found
to possess a continuum limit \cite{KKMW}. 
The scaling behavior expected from this result
has been correctly reproduced 
by a numerical simulation \cite{TsudaYukawa}. 

\begin{figure}
{\centering \resizebox*{0.8\columnwidth}{!}{\includegraphics{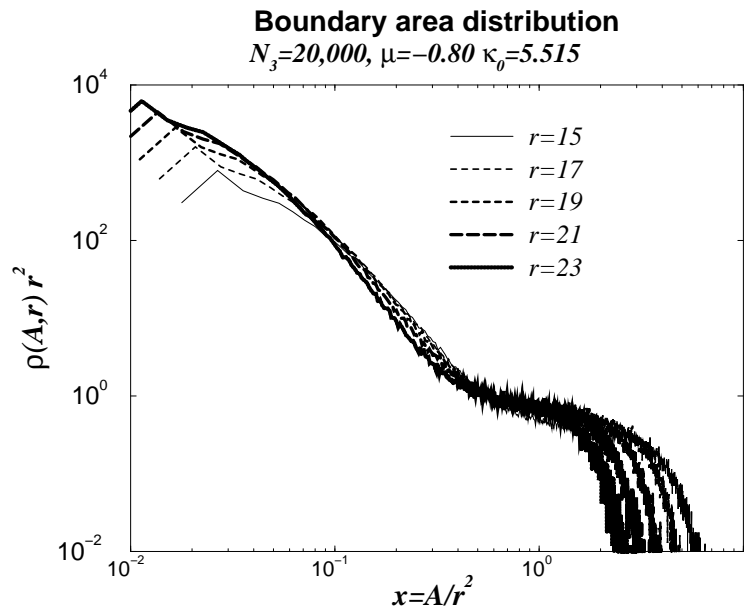}} \par}

\caption{Boundary area distribution \protect\( \rho (A,r)\protect \) 
at the end point of the first order phase transition
line with the system size \protect\( N_{3}=20,000\protect \).\label{fig:BAD}}
\end{figure}

In Fig. \ref{fig:BAD}, we plot the boundary area distribution 
at the end point of the first
order phase transition line as a function of \( x=A/r^{2} \). 
One can see a reasonable scaling
behavior. The power of \( r \) in the scaling variable \( x=A/r^{2} \) 

implies that the fractal
dimension is 3. We would like to remark that the scaling behavior we observe
is much better than the one that has been claimed to exist with the unmodified
action \cite{Hagura}. 

\section{Conclusions and discussions\label{Sec:Concl}}

We applied the multicanonical technique to the three dimensional dynamical
triangulation model. The microcanonical
inverse temperature served as a useful tool
to clarify the first order nature of the phase transition 
observed for the Einstein-Hilbert action and to 
investigate its fate in the expanded phase diagram
with a new coupling constant.

Our results about the modified action suggest 
that the first order phase transition
becomes second order at the end point, 
where one can  construct a continuum theory.
A natural question to be asked is what the continuum theory can be. 
One may suspect
that we cannot construct any continuum theory 
in three-dimensional quantum gravity,
since we have no physical degrees of freedom. 
This argument is too naive, however.
An example of field theory which has no physical degrees of freedom that is
still completely well-defined is two-dimensional Yang-Mills theory. 

The fact that we have to fine-tune two parameters 
to obtain the continuum limit
implies that there are two relevant operators around the fixed point. 
A possible
interpretation of the continuum theory is therefore the \( R^{2} \) gravity. 
The observation
that the fractal dimension at the fixed point is approximately three is also
suggestive of this interpretation. The fractal dimension we extracted, however,
is still preliminary, and the real value, which could be obtained by
increasing the system size
might well be larger. 

In Ref. \cite{Catterall}, the authors pointed out that
the system in the negative $\mu$ region might be suffered with
large finite size effects.
Taking this into account, they claimed that 
the line of first order phase transition continues to
$\mu = \mu ^ {c}$ and $\kappa_0 = \infty$ and
that the system is in the crumpled phase for all $\kappa_0$ when 
$\mu < \mu ^ {c}$.
Needless to say, simulations with larger system size
are necessary to discriminate between these possibilities.

In Ref. \cite{Progress}, 
we examined a different modified action in four-dimensional 
dynamical triangulation model. The added term was
\begin{equation}
\label{eq:SU}
S_{u}=u\sum _{v}\left[ o(v)-5 \right] ^{2}.
\end{equation}
We saw that the phase transition 
turned into a cross over even 
for a small positive \( u \) and 
the system was observed to be always in the branched polymer phase.
Let us write the term to be added as 
$ \sum _{v} f(o(v))$.
Note that a constant term and a linear term in $f(x)$ can be
absorbed by the redefinition of $\kappa_0$ and $\kappa_3$.
The quadratic term in (\ref{eq:SU}) might be too strong
since it dominates the behavior of $f(x)$ in the large $x$ region.

In Ref. \cite{Catterall}, another possibility
has been suggested that
the line of the first order phase transition ends at 
a point with {\it positive} $\mu$.
It is of interest to clarify this possibility
as well using the technique we used, although the simulation must be 
much harder due to the small acceptance rate \cite{Catterall}.

\bigskip

\noindent {\bf\Large Acknowledgments}

\bigskip

We would like to thank A. Fujitsu for helpful discussions and comments.
Most of the calculations in this paper has been done
on work stations in Aizu University.

\end{document}